\documentclass{amsart}
\usepackage{eurosym}
\usepackage{graphicx}
\usepackage[english]{babel}
\usepackage{multicol}
\usepackage{epstopdf}

\theoremstyle{definition}

\theoremstyle{remark}

\numberwithin{equation}{section}

\begin{document}
\title{Similarity Solutions for the Complex Burgers' Hierarchy}
\author{Amlan K Halder}
\address{Department of Mathematics, Pondicherry University, Kalapet,
India-605014}
\email{amlan.haldar@yahoo.com}
\thanks{AH expresses grateful thanks to UGC (India), NFSC, Award No.
F1-17.1/201718/RGNF-2017-18-SC-ORI-39488 for financial support and Late
Prof. K.M. Tamizhmani for the discussions which formed the basis of this work.%
}
\author{A. Paliathanasis}
\address{Instituto de Ciencias F\'{\i}sicas y Matem\'{a}ticas, Universidad
Austral de Chile, Valdivia, Chile\\
}
\address{Institute of Systems Science, Durban University of Technology, PO
Box 1334, Durban 4000, RSA\\
}
\email{anpaliat@phys.uoa.gr}
\thanks{AP acknowledges the financial support of FONDECYT grant no. 3160121.
}
\author{S. Rangasamy}
\address{Department of Mathematics, Shanmugha Arts Science Technology and
Research Academy, Thanjavur, India 613401}
\email{rsinuvasan@gmail.com}
\thanks{RS acknowledges Department of Science and Technology, Government of
India, FIST Programme SR/FST/MSI-107/2015.}
\author{PGL Leach}
\address{School of Mathematical Sciences, University of KwaZulu-Natal,
Durban, South Africa and\\
Institute of Systems Science, Durban University of Technology, Durban, South
Africa}
\email{Leachp@ukzn.ac.za}
\thanks{PGLL acknowledges the support of the National Research Foundation of
South Africa, the University of KwaZulu-Natal and the Durban University of
Technology and thanks the Department of Mathematics, Pondicherry University,
for gracious hospitality.}
\date{}
\dedicatory{}

\begin{abstract}
A detailed analysis of the invariant point transformations for the first
four partial differential equations which belong to the Complex Burgers'
Hierarchy is performed. Moreover, a detailed application of the reduction
process through the Lie point symmetries is presented while we construct
similarity solutions. We conclude that the differential equations of our
consideration are reduced to first-order equations such as the Abel, Riccati
and to a linearisable second-order differential equation by using similarity
transformations.
\end{abstract}

\maketitle









\vspace{1.5cc}

\section{Introduction}

Burgers' equation has been the centre of attraction for decades for its
diverse applications in different fields \cite{bur1,bur2,burg 06,burg 05}.
Its importance is the mathematical formulation for various subjects in
applied mathematics \cite{e1,e2,e3,burger3,e5}. In this paper we study the
symmetries of the Complex Burgers' Hierarchy \cite{burg 29}. The hierarchy
is given by the formula
\begin{equation}
u_{t}=t(L)P(iu_{x}e^{-i(u-\bar{u})}),  \label{1.1}
\end{equation}%
where $t(z)$ is an arbitrary entire function and the operators $P$ and $L$
are defined as \cite{burg 29}
\begin{eqnarray}
P(\beta (t,x)) &=&ie^{i(u-\bar{u})}\beta (t,x)\quad \mbox{\rm and}  \notag
\label{1.2} \\
L(\tau (t,x)) &=&i\tau _{x}+u_{x}\tau (t,x).  \notag
\end{eqnarray}

The dependent variables, $u$ and $\tau $, are functions of $t$ and $x$ of
complex type. The members of the hierarchy are obtained by setting $%
t(L)=L^{n} $, where $n=0,1,2,3...$. For $n=0,$ it leads to $u_{t}=-u_{x}$.
Subsequently, for higher values of $n$, the other members are obtained
eventually. For $n=1$ the second member of the hierarchy is
\begin{equation}
u_{t}=-u_{x}^{2}-iu_{xx}.  \label{1.3}
\end{equation}

This work focuses on the study of certain members of the Complex Burgers'
Hierarchy through Lie's approach. Lie symmetry analysis is a powerful method
for the study of nonlinear differential equations and there are many
applications of Lie's theory in different subjects of applied mathematics
\cite{aplie1,aplie2,aplie3,aplie4}. For instance for the classical Burgers
Equation (\ref{1.3}) the symmetry analysis has been performed in \cite%
{burger1}, while the symmetry analysis for the $2+1$ Burgers Equation is given
in \cite{burger2,burger3}.

The importance of the Lie symmetries is that they provide us with
differential invariants which can be used to reduce the order of
differential equations and to construct similarity solutions for the
original equation \cite{burg 20}. Hence in this work we study the algebraic
properties for the members of the complex Burgers' Hierarchy and we compare
the admitted Lie algebras and infer conclusions.

Furthermore we derive the travelling-wave solution for each of the members
by using the Lie invariants. Moreover the Lie point symmetries are applied
to determine travelling-wave solutions. For our analysis the Mathematica
package SYM was used \cite{burg 17,burg 18,burg 19}.

\section{The members of the Complex Burgers' Hierarchy}

We study the algebraic properties of the first four equations of the Complex
Burgers' Hierarchy. The first member of the Complex Burgers' Hierarchy is%
\begin{equation}
u_{t}+u_{x}=0~,~
\end{equation}%
where $u\left( t,x\right) $ is a complex function. By substitution of $%
u\left( t,x\right) =v\left( t,x\right) +iw\left( t,x\right) ,$ where $v,w$
are real functions, from the latter equation there follows the system%
\begin{eqnarray}
v_{t}+v_{x} &=&0,  \label{eq01a} \\
w_{t}+w_{x} &=&0.  \label{eq01b}
\end{eqnarray}

The second member of the Complex Burgers' Hierarchy is
\begin{eqnarray}
v_{t} &=&-v_{x}^{2}+w_{x}^{2}+w_{xx},  \label{eq02a} \\
w_{t} &=&-2v_{x}w_{x}-v_{xx}  \label{eq02b}
\end{eqnarray}%
when it is reduced to its real and imaginary parts. In terms of real
functions the second member of the Burgers' Hierarchy is well-known to be
linearisable by the Cole-Hopf transformation \cite{coleh}.

The third member of the Complex Burgers' Hierarchy is written in terms of
its components as
\begin{eqnarray}
v_{t} &=&-v_{x}^{3}+3v_{x}w_{x}^{2}+3w_{x}v_{xx}+3v_{x}w_{xx}+v_{xxx},
\label{eq03a} \\
w_{t} &=&-3v_{x}^{2}w_{x}+w_{x}^{3}-3v_{x}v_{xx}+3w_{x}w_{xx}+w_{xxx}.
\label{eq03b}
\end{eqnarray}%
The latter system is also called the complex Sharma-Tasso-Olver Equation
\cite{sh1,sh2,sh3}.

Finally the fourth member of the Complex Burgers' Hierarchy is
\begin{equation}
u_{t}=u_{x}^{4}+3u_{xx}^{2}+4u_{x}u_{xxx}+i(-3u_{x}u_{xx}-3u_{x}^{2}u_{xx}+u_{xxxx}),
\end{equation}%
for which the corresponding real and imaginary parts are
\begin{align}
v_{t}&=\begin{aligned}[t] &v_x^4 -6 v_x^2 w_x^2 +w_x^4 +3 w_x v_{xx}+6v_x
w_x v_{xx}+ 3 v_{xx}^2 +3 v_x w_{xx} +3 v_x^2 w_{xx}\nonumber\\ &-3 w_x^2
w_{xx}-3 w_{xx}^2 +4 v_x v_{xxx}-4 w_x w_{xxx}-w_{xxxx},\label{eq04a}\\
\end{aligned} \\
\end{align}
\mbox{\rm and}
\begin{align}
w_{t}&=\begin{aligned}[t] &4 v_x^3 w_x -4 v_x w_x^3 -3 v_x v_{xx} -3 v_x^2
v_{xx} +3 w_x^2 v_{xx} +3 w_x w_{xx} +6 v_x w_x w_{xx}+\nonumber\\ &6
v_{xx}w_{xx} + 4 w_x v_{xxx}+4 v_{x}w_{xxx} +v_{xxxx}.\label{eq04b}\\
\end{aligned} \\
\end{align}

For the set of four differential equations above, we apply Lie's theory and
we determine the point transformations under which the partial differential
equations are invariant.

\section{Lie symmetries and differential invariants}

For the convenience of the reader we briefly discuss the theory of Lie
symmetries of differential equations and the application of the differential
invariants for the construction of similarity solutions.

Let $\Phi $ be the map of an one-parameter point transformation such as
\begin{equation}
\Phi \left( u^{A}\left( t,x\right) \right) =u^{\prime A}\left( t,x\right)
\label{sv.10a}
\end{equation}%
$\ $with infinitesimal transformation ({$\varepsilon $ is the parameter of
smallness})%
\begin{eqnarray}
t^{\prime } &=&t+\varepsilon \xi ^{1}\left( t,x,u^{B}\right)  \label{sv.11}
\\
x^{\prime } &=&x+\varepsilon \xi ^{2}\left( t,x,u^{B}\right)  \label{sv.12}
\\
y^{\prime } &=&u^{A}+\varepsilon \eta ^{A}\left( t,x,u^{B}\right)
\label{sv.13}
\end{eqnarray}%
and generator
\begin{equation}
X=\frac{\partial t^{\prime }}{\partial \varepsilon }\partial _{t}+\frac{%
\partial x^{\prime }}{\partial \varepsilon }\partial _{x}+\frac{\partial
u^{A\prime }}{\partial \varepsilon }\partial _{u^{A}}  \label{sv.16}
\end{equation}%
in which $u^{A}\left( t,x\right) =\left( v\left( t,x\right) ,w\left(
t,x\right) \right) .$

Consider now that $u^{A}\left( t,x\right) $ is a solution of the partial
differential equation $\mathcal{H}\left(
u^{A},u_{,t}^{A},u_{,x}^{A}...\right) =0$. Then under the map $\Phi $
defined by (\ref{sv.10a}), function $u^{A\prime }\left( t^{\prime
},x^{\prime }\right) $ is also a solution of the differential equation $%
\mathcal{H}\left( u^{A},u_{,t}^{A},u_{,x}^{A}...\right) =0$ if and only if $%
\Phi \left( \mathcal{H}\left( u^{A},u_{,t}^{A},u_{,x}^{A}...\right) \right)
=0$, that is, the differential equation $\mathcal{H}\left(
u^{A},u_{,t}^{A},u_{,x}^{A}...\right) =0$ is invariant under the action of
the map, $\Phi $.

If this property be true, then the generator, $X,$ of the
infinitessimal transformation of the one-parameter point transformation, $%
\Phi $, \ is a Lie (point) symmetry of the differential equation \ $\mathcal{%
H}\left( u^{A},u_{,t}^{A},u_{,x}^{A}...\right) =0$. Mathematically that is
expressed as
\begin{equation}
X^{\left[ n\right] }\left( \mathcal{H}\right) =0,  \label{sv.17}
\end{equation}%
or equivalently
\begin{equation}
X^{\left[ n\right] }\left( \mathcal{H}\right) =\psi \mathcal{H},~mod\left(
\mathcal{H}\right) =0,  \label{sv.18}
\end{equation}%
where $X^{\left[ n\right] }$ denotes the $n-$prolongation/extension of the
symmetry vector in the space of variables $\left\{
t,x,u^{A},u_{,t}^{A},u_{,x}^{A},...\right\} $. The symmetry condition (\ref%
{sv.17}) provides a set of differential equations the solution of which
provides the generator of the infinitesimal transformation, (\ref{sv.16}).

The importance of the existence of a Lie symmetry for a partial differential
equation is that from the associated Lagrange's system,%
\begin{equation}
\frac{dt}{\xi ^{1}}=\frac{dx}{\xi ^{2}}=\frac{du^{A}}{\eta ^{A}},
\end{equation}%
zeroth-order invariants,~$U^{\left[ 0\right] }\left( t,x,u^{A}\right) $, can
be determined which can be used to reduce the number of the independent
variables of the differential equation and lead to the construction of
similarity solutions.

\subsection{Lie symmetries for the first member of the Complex Burgers'
Hierarchy.\newline
}

We continue by presenting the Lie point symmetries for the set of equations (%
\ref{eq01a})-(\ref{eq01b}). Specifically this system admits the infinite
number of symmetries
\begin{equation}
\Gamma =\begin{aligned}[t] &\xi^{1}(x, t, v, w)\partial_x +(c_{c}( -t +
x,v,w) +\xi^{1}(x, t, v, w))\partial_t\nonumber\\ &+c_{a}(-t +
x,v,w)\partial_v +c_{b}(v, w, -t + x)\partial_w\nonumber\\ \end{aligned}.
\label{2.3}
\end{equation}

The system of differential equations (\ref{eq01a})-(\ref{eq01b}) is
well-known to admit a travelling-wave solution. That family of solutions can
be easily derived by considering that functions $\xi ^{1}(t,x,v,w)$ and%
\newline
$c_{c}(x-t,v,w)$ are constants. That leads to the differential invariants%
\begin{equation}
s=x-ct~,~~~f(s)=v(t,x)~,~g(s)=w(t,x),  \label{2.33}
\end{equation}%
where now the system of differential equations, (\ref{eq01a})-(\ref{eq01b}),
is simplified to
\begin{equation}
\frac{d}{ds}f\left( s\right) =0~,~\frac{d}{ds}g\left( s\right) =0
\end{equation}%
with similarity solution $f\left( s\right) =f_{0}~,~g=g_{0}.$

We continue with the determination of the Lie point symmetries for the
system of differential equations, (\ref{eq02a})-(\ref{eq02b}).

\subsection{Lie symmetries for the second member of the Complex Burgers'
Hierarchy.\newline
}

The system of differential equations, (\ref{eq02a})-(\ref{eq02b}), admits
the following generic symmetry vector%
\begin{equation}
\Gamma =\begin{aligned}[t] &\bigg(A_{0}+A_{1}t+A_{2}
t^2\bigg)\partial_t+\bigg(A_{3}+A_{4}t+\frac{(A_{1}+2
A_{2}t)x}{2}\bigg)\partial_x+\nonumber\\
&\bigg(A_{5}-\frac{A_{2}t}{2}+e^{-w}
\cos(v)a(t,x)-e^{-w}b(t,x)\sin(v)\bigg)\partial_w+\nonumber\\
&\bigg(A_{6}+\frac{(2 A_{4} x+A_{2}
x^2)}{4}-e^{-w}\cos(v)b(t,x)-e^{-w}a(t,x)\sin(v)\bigg)\partial_v,\nonumber\\
\end{aligned}  \label{2.5}
\end{equation}%
where $A_{0--6}~$ are arbitrary constants, $a\left( t,x\right) $ and $b\left(
t,x\right) $ are functions which satisfy the linear constant coefficient $%
(1+1)$ evolution equations
\begin{equation}
a_{t}-a_{xx}=0~,~b_{t}-b_{xx}=0.
\end{equation}

From the latter it is clear that the system, (\ref{eq02a})-(\ref{eq02b}),
admits seven plus infinity Lie symmetry vectors. \ The seven vector fields
corresponds to the seven arbitrary constants $A_{0-6}$ and are%
\begin{eqnarray*}
\Gamma _{1a} &=&\partial _{t}, \\
\Gamma _{2a} &=&t\partial _{t}+\frac{x}{2}\partial _{x}, \\
\Gamma _{3a} &=&t^{2}\partial _{t}+tx\partial _{x}+\frac{x^{2}}{4}\partial
_{v}-\frac{t}{2}\partial _{w}, \\
\Gamma _{4a} &=&\partial _{x}, \\
\Gamma _{5a} &=&t\partial _{x}+\frac{x}{2}\partial _{v}, \\
\Gamma _{6a} &=&\partial _{v}, \\
\Gamma _{7a} &=&\partial _{w}.
\end{eqnarray*}


The Lie Brackets between the symmetries are
\begin{equation*}
\begin{split}
\lbrack \Gamma _{1a},\Gamma _{5a}]& =\Gamma _{1a} \\
\lbrack \Gamma _{1a},\Gamma _{6a}]& =\Gamma _{2a} \\
\lbrack \Gamma _{1a},\Gamma _{7a}]& =2\Gamma _{5a}-\frac{\Gamma _{4a}}{2} \\
\lbrack \Gamma _{2a},\Gamma _{5a}]& =\frac{\Gamma _{2a}}{2}
\end{split}%
\qquad
\begin{split}
\lbrack \Gamma _{2a},\Gamma _{6a}]& =\frac{\Gamma _{3a}}{2} \\
\lbrack \Gamma _{2a},\Gamma _{7a}]& =\Gamma _{6a} \\
\lbrack \Gamma _{5a},\Gamma _{6a}]& =\frac{\Gamma _{6a}}{2} \\
\lbrack \Gamma _{5a},\Gamma _{7a}]& =\Gamma _{7a}
\end{split}%
\end{equation*}%
from which we can infer that the symmetry vectors form the Lie algebra $%
A_{3,5}^{a}\oplus 2A_{1}.$ Hence the admitted Lie algebra for the system (%
\ref{eq02a})-(\ref{eq02b}) is
\begin{equation}
A_{3,5}^{a}\oplus 2A_{1}\oplus 2A_{1\infty }.
\end{equation}

\subsection{Lie symmetries for the third member of the complex Burgers'
Hierarchy.\newline
}

As far as concerns the Lie symmetries for the third member of the complex
Burgers' Hierarchy, i.e. system (\ref{eq03a})-(\ref{eq03b}), we find the
generic symmetry%
\begin{equation*}
\Gamma =\begin{aligned}[t] &\bigg(B_{0} + B_{1} t\bigg)\partial_t +
\bigg(B_{2} + \frac{B_{1} x}{3}\bigg)\partial_x +\nonumber\\ &\bigg(B_{5} -
\frac{B_{4} \cos(2 v)}{2} -e^{-w}(-d(t,x) \cos(v) + c(t,x)
\sin(v))+\frac{B_{3} \sin(2 v)}{2}\bigg)\partial_w +\nonumber\\ &\bigg(B_{6}
- e^{-w} (c(t,x) \cos(v) + d(t,x) \sin(v)) +\frac{B_{3} \cos(2 v)}{2} +
B_{4} \sin(2 v)\bigg)\partial_v\nonumber\\ \end{aligned},
\end{equation*}%
where $B_{0-7}$ are arbitrary constants, $c(t,x)$ and $d(t,x)$ satisfy the
linear evolution equations
\begin{eqnarray}
c_{t}-c_{xxx} &=&0\quad \mbox{\rm and}  \label{2.10} \\
d_{t}-d_{xxx} &=&0.
\end{eqnarray}%
From the general symmetry, we can write the seven vector fields which are%
\begin{eqnarray*}
\Gamma _{1b} &=&\partial _{t}, \\
\Gamma _{2b} &=&\partial _{t}+\frac{x}{3}\partial _{x}, \\
\Gamma _{3b} &=&\partial _{x}, \\
\Gamma _{4b} &=&\partial _{w}, \\
\Gamma _{5b} &=&\frac{\sin {2v}}{2}\partial _{v}-\frac{\cos {2v}}{2}\partial
_{w}, \\
\Gamma _{6b} &=&\frac{\cos {2v}}{2}\partial _{v}+\frac{\sin {2v}}{2}\partial
_{w}, \\
\Gamma _{7b} &=&\partial _{v}
\end{eqnarray*}%
for which the nonzero Lie Brackets are
\begin{equation*}
\begin{split}
\lbrack \Gamma _{1b},\Gamma _{3b}]& =\Gamma _{1b} \\
\lbrack \Gamma _{2b},\Gamma _{3b}]& =\frac{\Gamma _{2b}}{3}
\end{split}%
\qquad
\begin{split}
\lbrack \Gamma _{5b},\Gamma _{6b}]& =-\frac{\Gamma _{7b}}{2} \\
\lbrack \Gamma _{5b},\Gamma _{7b}]& =-2\Gamma _{6b}.
\end{split}%
\end{equation*}%
Hence, the Lie point symmetries for the third member of the complex Burgers'
hierarchy form the $A_{3,4}^{a}\oplus 3A_{1}\oplus 2A_{1\infty }$ Lie
algebra.

\subsection{Lie symmetries for the fourth member of the Complex Burgers'
Hierarchy.\newline
}

Finally, for the fourth member of the Complex Burgers' Hierarchy we find
that the system, (\ref{eq04a})-(\ref{eq04b}), admits only four Lie symmetry
vectors,%
\begin{eqnarray*}
\Gamma _{1c} &=&\partial _{t},~\Gamma _{2c}=\partial _{v}, \\
\Gamma _{3c} &=&\partial _{w},~\Gamma _{4c}=\partial _{x}
\end{eqnarray*}%
which constitute the Lie algebra $4A_{1}$ under the operation of taking the
Lie Bracket.

We continue our analysis with the application of the symmetry vectors to
reduce the system of partial differential equations and to find possible
similarity solutions. The reduction process is studied for the second, the
third and the fourth members of the hierarchy and more specifically we focus
on the travelling-wave solutions.\textbf{\ The reason that we choose to
perform the reduction by searching for travelling-wave solutions is because
that it is the only common reduction among all the four-members of the
hierarchy that we studied. With such an analysis we are able to compare the
travelling-wave solutions as we move to the higher-order members of the hierarchy.%
}

\section{Travelling-wave Solutions}

\subsection{Reduction process for the second member of the Complex Burgers'
Hierarchy.\newline
}

Consider the vector fields, $\Gamma _{1a}+c\Gamma _{4a}$, which are
symmetries of the system, (\ref{eq02a})-(\ref{eq02b}), and $c$ is a constant
which, as we see below, corresponds to the \textquotedblleft speed" of the
travelling-wave solution. The similarity variables, i.e. Lie invariants are given in (\ref%
{2.33}).

In the new variables equations~(\ref{eq02a})-(\ref{eq02b}) reduce to a
system of two second-order ordinary differential equations, namely,
\begin{eqnarray}
g^{\prime \prime }(s)-f^{\prime 2}+g^{\prime 2}+cf^{\prime }(s) &=&0,
\label{3.2a} \\
f^{\prime \prime }(s)+2f^{\prime }(s)g^{\prime }(s)-cg^{\prime }(s) &=&0.
\label{3.2b}
\end{eqnarray}

This system of ordinary differential equations admits a twelve-dimensional
algebra comprised of the vector fields
\begin{align*}
\Gamma _{1d}& =\begin{aligned}[t] &\partial_s\nonumber\\ \end{aligned} \\
\Gamma _{2d}& =\begin{aligned}[t] &\bigg(\frac{\cos{2 f} \cos{c
s}}{2}+\frac{\sin{2 f} \sin{c s}}{2}\bigg)\partial_f+\bigg(\frac{\cos{c s}
\sin{2 f}}{2}-\frac{\cos{2 f} \sin{c s}}{2}\bigg)\partial_g\nonumber\\
\end{aligned} \\
\Gamma _{3d}& =\begin{aligned}[t] &\bigg(\frac{\cos{2f} \sin{cs}}{2}
-\frac{\cos{cs}\sin{2f}}{2}\bigg)\partial_f+\bigg(\frac{\cos{2 f} \cos{c
s}}{2}+ \frac{\sin{2 f} \sin{c s}}{2}\bigg)\partial_g\nonumber\\
\end{aligned} \\
\Gamma _{4d}& =\begin{aligned}[t] &\partial_f\nonumber\\ \end{aligned} \\
\Gamma _{5d}& =\begin{aligned}[t] &\partial_g\nonumber\\ \end{aligned} \\
\Gamma _{6d}& =\begin{aligned}[t] &- e^{-g} \cos{f}\partial_f - e^{-g}
\sin{f}\partial_g\nonumber\\ \end{aligned} \\
\Gamma _{7d}& =\begin{aligned}[t] &\bigg(-\frac{ e^{-g} \cos{f} \cos{c
s}}{c}- \frac{e^{-g} \sin{f} \sin{c s}}{c}\bigg)\partial_f-\bigg(\frac{
e^{-g} \cos{c s} \sin{f}}{c} +\frac{e^{-g} \cos{f} \sin{c
s}}{c}\bigg)\partial_g\nonumber\\ \end{aligned} \\
\Gamma _{8d}& =\begin{aligned}[t] &\bigg(\frac{ e^{-g} \cos{c s} \sin{f}}{c}
- \frac{e^{-g} \cos{f} \sin{c s}}{c}\bigg)\partial_f+\bigg(-\frac{ e^{-g}
\cos{f} \cos{c s}}{c}- \frac{e^{-g} \sin{f} \sin{c
s}}{c}\bigg)\partial_g\nonumber\\ \end{aligned} \\
\Gamma _{9d}& =\begin{aligned}[t] & -e^{-g} \sin{f}\partial_f+e^{-g}
\cos{f}\partial_g\nonumber\\ \end{aligned} \\
\Gamma _{10d}& =\begin{aligned}[t] &e^{g} \cos{f}\partial_s+\frac{c e^{g}
\cos{f}}{2}\partial_f - \frac{ c e^{g}\sin{f}}{2}\partial_g\nonumber\\
\end{aligned} \\
\Gamma _{11d}& =\begin{aligned}[t] &\bigg(\frac{e^{g} \sin{f} \sin{c s}}{c}
+ \frac{e^{g} \cos{f} \cos{c s}}{c}\bigg)\partial_s+\bigg(\frac{
e^{g}\cos{f} \cos{c s}}{2}+ \frac{ e^{g} \sin{f} \sin{c
s}}{2}\bigg)\partial_f+\nonumber\\ &\bigg(\frac{ e^{g} \cos{c s}
\sin{f}}{2}- \frac{ e^{g} \cos{f} \sin{c s}}{2}\bigg)\partial_g\nonumber\\
\end{aligned} \\
\Gamma _{12d}& =\begin{aligned}[t] &-\bigg(\frac{e^{g} \cos{c s}
\sin{f}}{c}+\frac{e^{g} \cos{f} \sin{c s}}{c}\bigg)\partial_s-\bigg(\frac{
e^{g} \cos{c s} \sin{f}}{2}+\frac{ e^{g} \cos{f} \sin{c
s}}{2}\bigg)\partial_f+\nonumber\\ &\bigg(\frac{ e^{g} \cos{f} \cos{c
s}}{2}+\frac{ e^{g} \sin{f} \sin{c s}}{2}\bigg)\partial_g.\nonumber\\
\end{aligned}
\end{align*}

Easily the system, (\ref{3.2a})-(\ref{3.2b}), is reduced to the following
first-order equations%
\begin{eqnarray}
G^{\prime }\left( s\right) +cF\left( s\right) +G^{2}\left( s\right)
-F^{2}\left( s\right) &=&0  \label{3.3a} \\
F^{\prime }\left( s\right) +2F\left( s\right) G\left( s\right) -cG\left(
s\right) &=&0,  \label{3.3b}
\end{eqnarray}%
where $G\left( s\right) =g^{\prime }\left( s\right) $ and $F\left( s\right)
=f^{\prime }\left( s\right) $. Easily the solution of the latter system can
be written in closed form as
\begin{equation}
F\left( s\right) =\frac{c}{2}~,~G\left( s\right) =-\frac{c}{2}\tan \left(
\frac{c}{2}\left( s-s_{0}\right) \right)
\end{equation}%
or%
\begin{equation}
G\left( s\right) =-\frac{F^{\prime }}{2F-c}~,~F\left( s\right) =\frac{c}{2}%
\frac{F_{0}\left( e^{-2ics}-F_{1}c\right) ^{2}-16c^{2}-8cF_{0}e^{-ics}}{%
F_{0}\left( e^{-2ics}-F_{1}c\right) ^{2}-16c^{2}}.  \label{s11}
\end{equation}

The corresponding behaviour of the functions $F\left( s\right) ~$and $%
G\left( s\right) $, for various values of $F_{1}$, is plotted in Fig. \ref%
{fig.1} in which we can observe the existence of wave solutions.
\begin{figure}[h]
\centering
\includegraphics[width=0.8\textwidth]{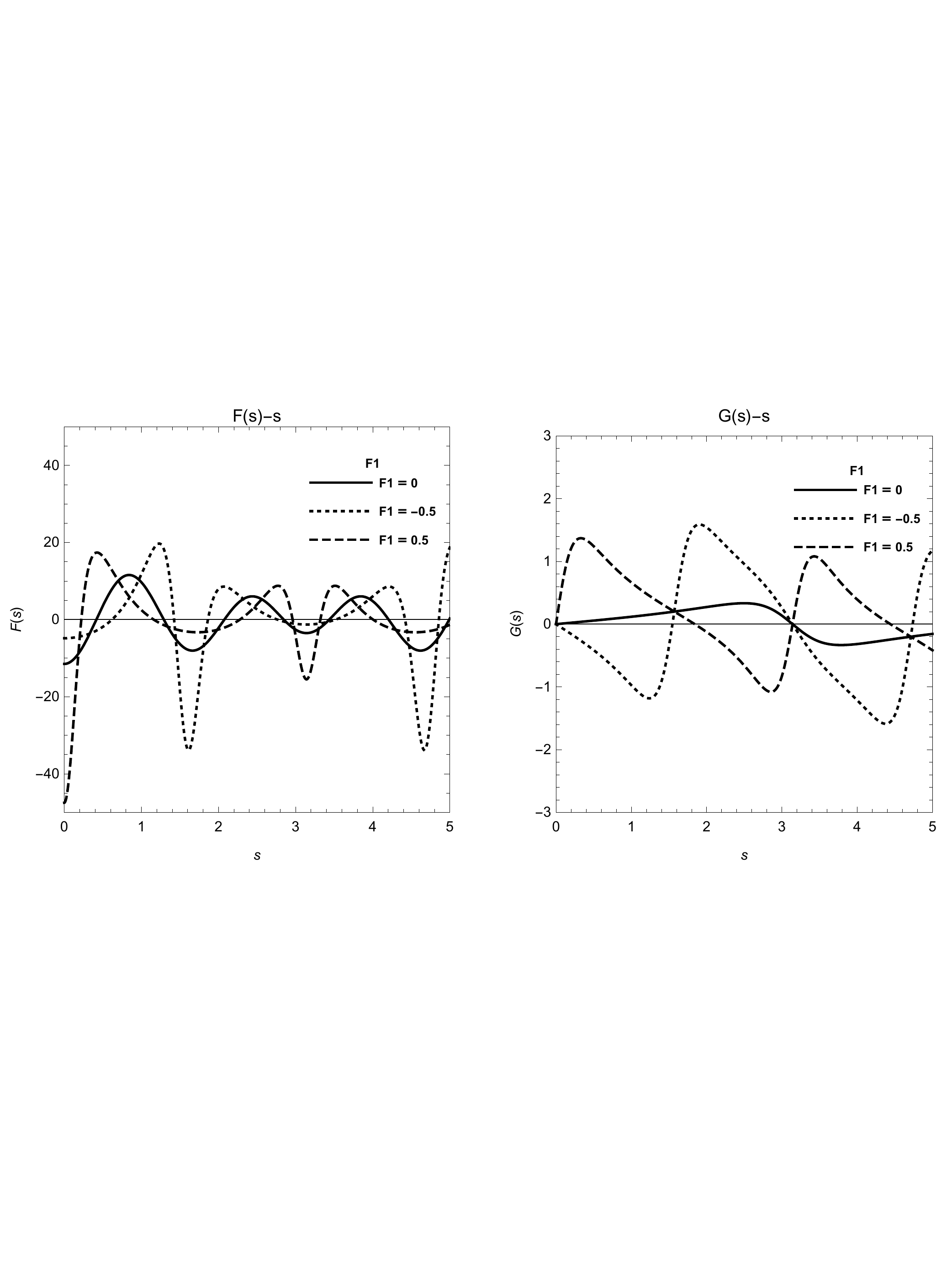}
\caption{Qualitative evolution for the solution (\protect\ref{s11}) for
different values of the constant of integration constant, $F_{1}$. Left figures is for
function $F\left( s\right) $ while right figure is for function $G\left(
s\right) $. }
\label{fig.1}
\end{figure}

As it is evident from the figure, less turbulence prevails for $F_{1}$ at $0$%
, as compared to other values. It is important to mention that equation (\ref%
{3.3a}) with use of (\ref{3.3b}) can be written as a second-order
differential equation,
\begin{equation}
\left( F^{\prime \prime }+\frac{1}{\left( 2F-c\right) }3F^{\prime 2}\right)
-F\left( F-c\right) \left( 2F-c\right) =0,
\end{equation}%
which is invariant under the elements of the $sl(2,R)$ Lie algebra. This
means that the equation can be easily transformed to the Ermakov-Pinney
Equation. We continue with the third member of the hierarchy.

\subsection{Reduction process for the third member of the Complex Burgers'
Hierarchy.\newline
}

The travelling-wave solution for the system, (\ref{eq03a})-(\ref{eq03b}),
with respect to the similarity variables for $\Gamma _{1b}+c\Gamma _{3b}$ is
of form of (\ref{2.33}). The reduced equations are
\begin{eqnarray}
f^{\prime \prime \prime }(s)+cf^{\prime }(s)-f^{\prime 3}+3f^{\prime
}(s)g^{\prime 2}+3g^{\prime }(s)f^{\prime \prime }(s)+3f^{\prime
}(s)g^{\prime \prime }(s) &=&0  \label{3.20} \\
g^{\prime \prime \prime }(s)+cg^{\prime }(s)-3f^{\prime 2}g^{\prime
}(s)+g^{\prime 3}-3f^{\prime }(s)f^{\prime \prime }(s)+3g^{\prime
}(s)g^{\prime \prime }(s) &=&0.  \label{3.20b}
\end{eqnarray}%
This system is invariant under the action of a five-dimensional Lie algebra
comprising the following vector fields
\begin{eqnarray}
\Gamma _{1f} &=&\partial _{f}  \notag  \label{3.22} \\
\Gamma _{2f} &=&\partial _{g}  \notag \\
\Gamma _{3f} &=&-\bigg(\frac{\sin {\sqrt{c}s}}{\sqrt{c}}\bigg)\partial
_{s}-\cos {\sqrt{c}s}\partial _{g}  \notag \\
\Gamma _{4f} &=&-\bigg(\frac{(\cos {\sqrt{c}s})}{\sqrt{c}}\bigg)\partial
_{s}+\sin {\sqrt{c}s}\partial _{g}  \notag \\
\Gamma _{5f} &=&\partial _{s}  \notag
\end{eqnarray}%
and nonzero Lie Brackets,
\begin{equation}
\lbrack \Gamma _{3f},\Gamma _{4f}]=-\frac{\Gamma _{5f}}{\sqrt{c}}~,~[\Gamma
_{3f},\Gamma _{5f}]=-\sqrt{c}\Gamma _{4f}~~,~[\Gamma _{4f},\Gamma _{5f}]=%
\sqrt{c}\Gamma _{3f},  \notag
\end{equation}%
that is, the vector fields, $\Gamma _{f}$, form the $2A_{1}\oplus _{s}so(2,1)~$%
Lie algebra.

The application of the autonomous symmetry $\partial _{s}$ in the system (%
\ref{3.20})-(\ref{3.20b}) leads us also to the autonomous system of
second-order differential equations,%
\begin{eqnarray}
F^{\prime \prime }-cF-F^{3}+3FG^{2}+3GF^{\prime }+3FG^{\prime }&=&0
\label{3.22a} \\
G^{\prime \prime }-cG-3F^{2}G+G^{3}-3FF^{\prime }+3GG^{\prime }&=&0,
\label{3.22b}
\end{eqnarray}%
where, as above, $G\left( s\right) =g^{\prime }\left( s\right) $ and $F\left(
s\right) =f^{\prime }\left( s\right) .~\ $From the Lie symmetries $\Gamma
_{1f}$ and $\Gamma _{3f}$ of the system (\ref{3.22a})-(\ref{3.22b}) we can
define the particular solution~ and
\begin{equation}
F\left( s\right) =0~,~G\left( s\right) =\sqrt{c}\frac{\sin \left( \sqrt{c}%
s\right) -G_{0}\cos \left( \sqrt{c}s\right) }{G_{0}\sin \left( \sqrt{c}%
s\right) +\cos \left( \sqrt{c}s\right) +G_{1}}.
\end{equation}%
By using the symmetry $\Gamma _{2f}$ we conclude that%
\begin{equation}
G\left( s\right) =0~,~F\left( s\right) =F_{0}sn\left( s,a_{1}\right)
\end{equation}%
where $sn\left( s,a_{1}\right) $ is the Jacobi elliptic function and $%
a_{0}\left( s\right) =\frac{\sqrt{2c}\left( 2a_{1}+\sqrt{4c-2}s\right) }{2%
\sqrt{a_{1}^{2}+2c-1}}$. In a similar way we can construct similarity
solutions by using the combination of the Lie symmetries $\Gamma _{1f}+\beta
\Gamma _{2f}$.


\subsection{Reduction process for the fourth member of the Complex Burgers'
Hierarchy.\newline
}

The reduced equations with respect to the $\Gamma _{1c}+c\Gamma _{4c},$ as
mentioned above, are
\begin{eqnarray}
g^{\prime \prime \prime \prime }(s)&=\begin{aligned}[t] &f'(s)^4 -6 f'(s)^2
g'(s)^2 + g'(s)^4 + 3 g'(s) f''(s) + 6 f'(s) g'(s) f''(s) +3 f''(s)^2 +3
f'(s) g''(s) +\nonumber\\ &3 f'(s)^2 g''(s) -3 g'(s)^2 g''(s) -3 g''(s)^2 +4
f'(s) f'''(s) +4 g'(s)
g'''(s)+cf^{\prime}(s),\label{3.47a}\nonumber\\\end{aligned} \\
f^{\prime \prime \prime \prime }(s)&=\begin{aligned}[t] &4 f'(s) g'(s)^3-4
f'(s)^3 g'(s)+3 f'(s) f''(s) +3 f'(s)^2 f''(s) -3 g'(s)^2 f''(s) -3 g'(s)
g''(s) -\nonumber\\ &6 f'(s) g'(s)g''(s) -6 f''(s) g''(s) -4 g'(s) f'''(s)+4
f'(s) g'''(s)-cg^{\prime}(s).\label{3.47b}.\nonumber\\\end{aligned}
\end{eqnarray}%
The Lie-Point symmetries of the resulting system are
\begin{equation*}
\Gamma _{1j}=\partial _{s}~,~\Gamma _{2j}=\partial _{f}~,~\Gamma
_{3j}=\partial _{g}.
\end{equation*}

The application of the Lie symmetry, $\partial _{s}$, leads us again to an
autonomous third-order dynamical system with only two symmetries, the $%
\Gamma _{2j}$ and $\Gamma _{3j}.$ From these symmetry vectors the only
possible solution that we can get is
\begin{equation}
g\left( s\right) =g_{0}~,~f\left( s\right) =f_{1}s+f_{0}.
\end{equation}%
This is a particular real solution. However, it is not a travelling-wave
solution. Hence, in order to study the existence of solutions we should
generalised the context of symmetries to the case of nonpoint symmetries or
use other methods for solving nonlinear differential equations.

\section{Conclusions}

This work focused on the study of the algebraic properties of the
differential equations which belong to the Complex Burgers' hierarchy. More
specifically, we studied the Lie point symmetries for the first four
equations of the Complex Burgers' Hierarchy. We found that the first member
of the hierarchy is invariant under an infinite number of symmetries. The
second member of the hierarchy is invariant under the group of
transformations with generators the elements of the $A_{3,5}^{a}\oplus
2A_{1}\oplus 2A_{1\infty }$, which is comprised of seven plus two times
infinity symmetries. The third member of the hierarchy is invariant under $%
A_{3,4}^{a}\oplus 3A_{1}\oplus 2A_{1\infty }$. On the other hand, the fourth
member of the hierarchy is invariant under the four-dimensional Lie algebra $%
4A_{1}$.

From the symmetry analysis it is clear that the differential equations which
belong to the first three members of the Complex Burgers' hierarchy can be
linearised because the infinite number of point symmetries exists (vice versa). However, we
cannot reach a similar conclusion for the fourth member of the hierarchy, at
least as far as concerns point transformations. However, it is well-known
that the Burgers' hierarchy is linearised by the Cole-Hopf transformation
\cite{burg 29}.

As far as concerns the number of admitted Lie point symmetries, someone
may expect a common feature among the different members of the
hierarchy. However, that is not true. We observe that as we proceed through
the hierarchy, the number of symmetries decreases. The only common
symmetries are the time and space translation, $\left\{ \partial
_{t},~\partial _{x}\right\} $, which of course exist because the
differential equations are autonomous and homogeneous. The linear
combination of these two symmetries forms the $2A_{1}$ Lie algebra and
provides the similarity variables for the travelling-wave solutions.

We applied these two symmetries for all the members of the hierarchy of our
study and we reduced the systems of partial differential equations to
systems of ordinary differential equations. For these systems we determined
the Lie point symmetries and we proceeded with the further reduction. We
conclude that travelling-wave solutions can be determined explicitly by the
use of Lie point symmetries only for the first, second and third members of
the Complex Burgers' Hierarchy.

From our analysis it is clear that someone should generalise the context of
symmetries to nonpoint symmetries in order to study higher members of the
hierarchy and to determine analytic and exact solutions. Such an analysis is a
subject of a further study.

\end{document}